\documentclass[
pra,twocolumn,showpacs,preprintnumbers,amsmath,amssymb]{revtex4}

\usepackage{graphicx}
\usepackage{amssymb}
\usepackage{amsmath}
\usepackage{dsfont}
\usepackage{bm}
\usepackage{mathrsfs}
\usepackage{times}

\begin{document}
\title{Electromagnetic energy-momentum in dispersive media}
\author{T.\ G.\ Philbin}

\affiliation{School of Physics and Astronomy, University of St Andrews,
North Haugh, St Andrews, Fife, KY16 9SS, Scotland, United Kingdom}

\begin{abstract}
The standard derivations of electromagnetic energy and momentum in media take Maxwell's equations as the starting point. It is well known that for dispersive media this approach does not directly yield exact expressions for the energy and momentum densities. Although Maxwell's equations fully describe electromagnetic fields, the general approach to conserved quantities in field theory is not based on the field equations, but rather on the action. Here an action principle for macroscopic electromagnetism in dispersive, lossless media is used to derive the exact conserved energy-momentum tensor. The time-averaged energy density reduces to Brillouin's simple formula when the fields are monochromatic. The momentum density is not given by the familiar Minkowski expression $\mathbf{D}\times\mathbf{B}$, even for time-averaged monochromatic fields. The results are unaffected by the debate over momentum balance in light-matter interactions.
\end{abstract}
\pacs{03.50.De,42.25.Bs,42.50.Wk}

\maketitle

[{\it Note: This arXiv version incorporates the Erratum to the published version~\cite{paper}.}]

\section{Introduction} \label{sec:intro}
Momentum balance in light-matter interactions is a subtle subject that has been debated for over a century. Discussions of the issues involved, together with references to much of the relevant literature, can be found in~\cite{bre79,leo06,pfe07,bar10a,bar10b}. This paper makes no attempt to contribute to the momentum-balance debate, in that there will be no investigation of the transfer of momentum (or energy) from light to matter or vice versa. Instead we isolate a related problem, but one that is mathematically well-posed and therefore capable of an exact and unambiguous solution. The problem solved here is: what is the conserved energy-momentum tensor of light propagating in a homogeneous, dispersive, lossless magnetodielectric medium? This is a well-posed problem of macroscopic electromagnetism, but it appears not to have been fully addressed in the existing literature on electromagnetic energy-momentum, where the primary focus has been on momentum transfer between light and matter. Although the solution is an example of Noether's theorem, the presence of dispersion means that the well-known expression for the energy-momentum tensor of a field theory in terms of its Lagrangian density is not valid in this case. In fact the result cannot be derived in any systematic way from Maxwell's equations or a Lagrangian density. Dispersive macroscopic electromagnetism must be formulated in terms of an action principle, and the Noether theorem for space-time translation symmetry must be derived from the action, leading to an exact conserved energy-momentum tensor. The problem in fact presents a significant calculational challenge, which may explain its absence from the momentum-balance literature. By isolating this problem from the conundrum of momentum transfer, we show that the issue of electromagnetic energy-momentum in dispersive media is amenable to some mathematically exact and unambiguous statements. 

The restriction to lossless media in this paper requires comment on the physical significance of the results. A complete treatment of macroscopic electromagnetism at all frequencies must include the absorption in the materials that is necessarily strong in some frequency ranges. Mathematically, this dissipation is a consequence of the restriction to retarded solutions of Maxwell's equations, which leads to the Kramers-Kronig relations~\cite{jac,LLcm}. As is well known~\cite{LLcm,bar76,ros10}, quantifying electromagnetic energy-momentum in dissipative media is problematic, and in any case such energy-momentum will of course not be conserved. In optics, on the other hand, one often deals with limited frequency ranges where losses are negligible, but where material dispersion cannot be ignored. An example of the utmost experimental and practical importance is fiber optics~\cite{agr}, where in many circumstances absorption can be ignored not only in the entire visible range, but also into the infrared and ultraviolet. In such a large frequency range dispersion plays a crucial role in light propagation and cannot be neglected~\cite{agr}. The question of how much conserved energy-momentum is being transported by light has an unambiguous answer in these circumstances, and the result is experimentally significant. For example, a light beam encountering an intense pulse can be frequency shifted through non-linearity of the medium, and the amount of electromagnetic energy that is frequency shifted depends, among other factors, on the energy-momentum tensor of the light beam. One example of such an experiment is described in~\cite{phi08}, where the light beam encountering the intense pulse was monochromatic; this represents the simplest case, but for a light beam with a complicated spectrum the general expression for the energy-momentum tensor is required to fully describe the behaviour of the beam. This exact energy-momentum tensor is derived here.

The results obtained below generalize a well-known approximation for electromagnetic energy in dispersive media. As is familiar from the textbook treatments~\cite{jac,LLcm}, the Poynting theorem that follows from Maxwell's equations does not directly lead to an expression for the electromagnetic energy density in the case of lossless, dispersive media. The standard procedure~\cite{jac,LLcm} is to make a restriction to quasi-monochromatic fields, in which case one can extract an approximate expression for the time-averaged energy density, due to Brillouin (Eq.\ (\ref{brill}) below). For general fields one is reduced to making a formal time integration of the Poynting theorem~\cite{bar76}. But as noted above, in the dispersive case one cannot derive Noether's theorem from a systematic manipulation of the dynamical equations of the theory, as is done in the Poynting theorem; instead, one must start with the action that underlies the equations. Noether's theorem guarantees the existence of an exact energy density for arbitrary fields in lossless dispersive media and this energy density is derived here.

In the case of electromagnetic momentum density, the dispersionless result is the familiar Minkowski expression~\cite{bre79,leo06,pfe07,bar10a,bar10b} $\mathbf{D}\times\mathbf{B}$.  We note again that this statement refers to the conserved electromagnetic energy-momentum tensor: the Minkowski momentum density  $\mathbf{D}\times\mathbf{B}$ unambiguously gives the conserved electromagnetic momentum of light propagating in a non-dispersive, lossless, homogeneous medium~\cite{bre79,leo06,pfe07,bar10a,bar10b}. It is shown below that in dispersive media the momentum density is more complicated, and is not given by Minkowski's formula. 

\section{The action}  \label{sec:act}
The results follow straightforwardly once an action principle is written for electromagnetism in dispersive, lossless media. For this purpose the dynamical variables must be taken to be the scalar potential $\phi$ and vector potential $\mathbf{A}$, defined by
\begin{equation} \label{pots}
\mathbf{E}=-\nabla\phi-\partial_t\mathbf{A}, \quad \mathbf{B}=\nabla\times\mathbf{A}.
\end{equation}
The relationship between fields in the time and frequency domains (the latter denoted by a tilde) is, for the example of the electric field,
\begin{equation}  \label{Efreq}
\begin{split}
\mathbf{E}(\mathbf{r},t)=&\frac{1}{2\pi}\int_0^\infty d\omega\left(\tilde{\mathbf{E}} (\mathbf{r},\omega)e^{-i\omega t}+\text{c.c}\right) \\
=&\frac{1}{2\pi}\int_{-\infty}^\infty d\omega\,\tilde{\mathbf{E}}(\mathbf{r},\omega)e^{-i\omega t},
\end{split}
\end{equation}
where the reality of $\mathbf{E}(\mathbf{r},t)$ implies $\tilde{\mathbf{E}}(\mathbf{r},-\omega)=\tilde{\mathbf{E}}^*(\mathbf{r},\omega)$. In the frequency domain we have
\begin{equation}
\tilde{\mathbf{D}}=\varepsilon_0\varepsilon(\mathbf{r},\omega)\tilde{\mathbf{E}}, \quad \tilde{\mathbf{H}}=\kappa_0\kappa(\mathbf{r},\omega)\tilde{\mathbf{B}},
\end{equation}
where $\varepsilon(\mathbf{r},\omega)$ is the relative permittivity of the (in general inhomogeneous) medium, $\kappa_0=\mu_0^{-1}$, and the relative permeability is $\mu(\mathbf{r},\omega)=\kappa(\mathbf{r},\omega)^{-1}$. As the medium can be assumed to be lossless in the frequency range of interest, $\varepsilon(\mathbf{r},\omega)$ and $\kappa(\mathbf{r},\omega)$ are real and even functions of $\omega$~\cite{LLcm}, and so they have the series expansions
\begin{equation} \label{matseries}
\varepsilon(\mathbf{r},\omega)=\sum_{n=0}^\infty\varepsilon_{2n}(\mathbf{r})\,\omega^{2n}, \ \kappa(\mathbf{r},\omega)=\sum_{n=0}^\infty \kappa_{2n}(\mathbf{r})\,\omega^{2n},
\end{equation}
where these equations define the coefficients $\varepsilon_{2n}(\mathbf{r})$ and $\kappa_{2n}(\mathbf{r})$. In practice, the series (\ref{matseries}) will represent a fit to the dispersion data of the material for the frequency range of interest; we take these series to be infinite, but the subsequent results also hold when they are finite series. Equations (\ref{Efreq})--(\ref{matseries}) show that in the time domain the Maxwell equations (with no free charges or currents) are
\begin{gather} 
\varepsilon_0\nabla\cdot[\varepsilon(\mathbf{r},i\partial_t)\mathbf{E}]=0,  \label{max1}  \\
\kappa_0\nabla\times[\kappa(\mathbf{r},i\partial_t)\mathbf{B}]=\varepsilon_0\varepsilon(\mathbf{r},i\partial_t)\partial_t\mathbf{E}, \label{max2}
\end{gather}
while the other two Maxwell equations are automatically satisfied because of (\ref{pots}). Note that series expansions of the form (\ref{matseries}) are standard in treating dispersion in frequency ranges where losses are negligible, for example in linear and nonlinear fiber optics~\cite{agr}. Moreover the transformation from the frequency to the time domain apparent in (\ref{max1})--(\ref{max2}), and the reverse transformation, are a standard part of numerical solution procedures for the propagation of wave packets with spectra within the frequency range where expansions (\ref{matseries}) are valid, for example in the split-step method~\cite{agr,ree03}. The action $\mathcal{S}[\phi,\mathbf{A}]$ for the potentials $\phi$ and $\mathbf{A}$ that gives the dynamical equations (\ref{max1})--(\ref{max2}) is
\begin{equation} \label{act}
\mathcal{S}=\int d^4x\frac{\kappa_0}{2}\left\{\frac{1}{c^2}\mathbf{E}\cdot[\varepsilon(\mathbf{r},i\partial_t)\mathbf{E}]-\mathbf{B}\cdot[\kappa(\mathbf{r},i\partial_t)\mathbf{B}]\right\}.
\end{equation}
Variation of $\phi$ in (\ref{act}) gives the Maxwell equation (\ref{max1}), while variation of $\mathbf{A}$ gives (\ref{max2}). 

The fact that only even-order time derivatives occur in  (\ref{max1})--(\ref{max2}) is essential to being able to write an action principle for these equations. For dissipative media, terms with odd-order time derivatives occur in (\ref{max1})--(\ref{max2}), arising from the imaginary parts of $\varepsilon(\mathbf{r},\omega)$ and $\kappa(\mathbf{r},\omega)$ which are odd functions of $\omega$~\cite{LLcm}; but the action (\ref{act}) would not generate these terms. In the action (\ref{act}), terms of the form $\mathbf{E}\cdot\partial_t^{2n+1}\mathbf{E}$, for example, would not contribute to the dynamical equations because their variation gives zero after integrations by parts: $\delta(\mathbf{E}\cdot\partial_t^{2n+1}\mathbf{E})=(\delta\mathbf{E})\cdot\partial_t^{2n+1}\mathbf{E}+\mathbf{E}\cdot\partial_t^{2n+1}\delta\mathbf{E}$, and the second term becomes minus the first upon integrations by parts. It must of course be impossible to write an
action in the dissipative case that is a functional only of $\phi$ and $\mathbf{A}$, since this would imply the existence of a conserved electromagnetic energy in lossy media (see below).

\section{Energy}  \label{sec:en}
The action (\ref{act}) is invariant under active time translations of the dynamical fields $\phi$ and $\mathbf{A}$, and this invariance implies, through Noether's theorem, the existence of a conserved quantity, the energy. The extraction of the conservation law from the action is a standard technique of field theory (see~\cite{wei}, for example). Even if one is unfamiliar with Noether's theorem, one can of course verify using the field equations  (\ref{max1})--(\ref{max2}) that the resulting conservation law does in fact hold. The theorem~\cite{wei} shows that if we make an active infinitesimal time translation $\phi(\mathbf{r},t)\rightarrow\phi(\mathbf{r},t+\zeta(\mathbf{r},t))$, $\mathbf{A}(\mathbf{r},t)\rightarrow\mathbf{A}(\mathbf{r},t+\zeta(\mathbf{r},t))$, but take the translation $\zeta(\mathbf{r},t)$ to vary in space and time, then the change in the action can be reduced to the form
\begin{equation} \label{actvart}
\delta\mathcal{S}=\int d^4x\left(\rho\,\partial_t\zeta+\mathbf{S}\cdot\nabla\zeta\right),
\end{equation}
where $\rho$ is the energy density and $\mathbf{S}$ is the energy flux, obeying the conservation law
\begin{equation} \label{cone}
\partial_t \rho+ \nabla \cdot\mathbf{S}=0.
\end{equation}
The variation $\delta\mathcal{S}$ will clearly be linear in the infinitesimal translation $\zeta(\mathbf{r},t)$, but various numbers of integrations by parts are required to achieve the form (\ref{actvart}) (surface terms produced by the integrations are to be dropped). Use must be made of the following identities, which hold for arbitrary functions $Y$ and $Z$, up to surface terms:
{\allowdisplaybreaks
\begin{align}
\int& d^4x\,Y\varepsilon(\mathbf{r},i\partial_t)(\zeta Z)
=\int d^4x{\Bigg[}\zeta Y\varepsilon(\mathbf{r},i\partial_t)Z \nonumber \\
&\left.-\sum_{n=1}^\infty\sum_{m=1}^{2n}(-1)^{n+m}\varepsilon_{2n}(\mathbf{r})\partial_t^{m-1}Y\partial_t^{2n-m}Z\partial_t\zeta\right], \label{sum1} \\
\int& d^4x\,Y\varepsilon(\mathbf{r},i\partial_t)\partial_t(\zeta Z)
=\int d^4x{\Bigg[}\zeta Y\varepsilon(\mathbf{r},i\partial_t)\partial_tZ \nonumber \\
&\left.+\sum_{n=0}^\infty\sum_{m=0}^{2n}(-1)^{n+m}\varepsilon_{2n}(\mathbf{r})\partial_t^{m}Y\partial_t^{2n-m}Z\partial_t\zeta\right]. \label{sum2}
\end{align}}%
These identities were found by first working with small values of $n$, where all the terms can be checked by hand, and then verifying the general expressions (\ref{sum1})--(\ref{sum2}) using Mathematica. Identity (\ref{sum1}) with $\varepsilon(\mathbf{r},i\partial_t)$ replaced by $\kappa(\mathbf{r},i\partial_t)$ must also be used. In this way one attains the form (\ref{actvart}) with
{\allowdisplaybreaks
\begin{align}
 \rho=&\frac{\kappa_0}{2}\left\{\frac{1}{c^2}(\nabla\phi-\partial_t\mathbf{A})\cdot[\varepsilon(\mathbf{r},i\partial_t)\mathbf{E}]+\mathbf{B}\cdot[\kappa(\mathbf{r},i\partial_t)\mathbf{B}]\right. \nonumber \\
&-\frac{1}{c^2}\sum_{n=1}^\infty\sum_{m=1}^{2n}(-1)^{n+m}\varepsilon_{2n}(\mathbf{r})\partial_t^{m-1}\mathbf{E}\cdot\partial_t^{2n-m+1}\mathbf{E}  \nonumber \\
&+\left.\sum_{n=1}^\infty\sum_{m=1}^{2n}(-1)^{n+m}\kappa_{2n}(\mathbf{r})\partial_t^{m-1}\mathbf{B}\cdot\partial_t^{2n-m+1}\mathbf{B}\right\},  \label{rho1} \\
\mathbf{S}=& -\varepsilon_0\partial_t\phi\,\varepsilon(\mathbf{r},i\partial_t)\mathbf{E}-\kappa_0\partial_t\mathbf{A}\times[\kappa(\mathbf{r},i\partial_t)\mathbf{B}]. \label{S1}  
\end{align}}%
It is straightforward to verify that (\ref{rho1}) and (\ref{S1}) obey the conservation law (\ref{cone}) when the fields satisfy Maxwell's equations (\ref{max1})--(\ref{max2}). As in the case of dispersionless media, and indeed vacuum, the energy density (\ref{rho1}) and flux (\ref{S1}) that directly emerge from Noether's theorem are not gauge invariant~\cite{jac,wei}. They are however equivalent to gauge-invariant quantities because they fail to be gauge invariant up to terms that \emph{identically} satisfy the conservation law (\ref{cone}). Specifically, the quantities
\begin{gather}
f^{i\ t}_{\ \,t}:= -\varepsilon_0\phi\,\varepsilon(\mathbf{r},i\partial_t)E^i =:-f^{t\ i}_{\ \,t},  \\
f^{j\ i}_{\ \,t}:=-\kappa_0\phi\,\kappa(\mathbf{r},i\partial_t)(\nabla^iA^j-\nabla^jA^i),
\end{gather}
identically satisfy
\begin{equation}  \label{fid}
\partial_t\nabla_if^{i\ t}_{\ \,t}+\nabla_i(\partial_tf^{t\ i}_{\ \,t}+\nabla_jf^{j\ i}_{\ \,t})=0.
\end{equation}
Comparing (\ref{fid}) with (\ref{cone}), we see that if $\nabla_if^{i\ t}_{\ \,t}$ is added to $\rho$, and $\partial_tf^{t\ i}_{\ \,t}+\nabla_jf^{j\ i}_{\ \,t}$ is added to $S^i$, then the conservation law (\ref{cone}) will still hold. Moreover, with use of Maxwell's equations (\ref{max1})--(\ref{max2}), the energy density and flux that result from these additions are gauge invariant and are given by
{\allowdisplaybreaks
\begin{align}
 \rho=&\frac{\kappa_0}{2}\left\{\frac{1}{c^2}\mathbf{E}\cdot[\varepsilon(\mathbf{r},i\partial_t)\mathbf{E}]+\mathbf{B}\cdot[\kappa(\mathbf{r},i\partial_t)\mathbf{B}]\right. \nonumber \\
&-\frac{1}{c^2}\sum_{n=1}^\infty\sum_{m=1}^{2n}(-1)^{n+m}\varepsilon_{2n}(\mathbf{r})\partial_t^{m-1}\mathbf{E}\cdot\partial_t^{2n-m+1}\mathbf{E}  \nonumber \\
&+\left.\sum_{n=1}^\infty\sum_{m=1}^{2n}(-1)^{n+m}\kappa_{2n}(\mathbf{r})\partial_t^{m-1}\mathbf{B}\cdot\partial_t^{2n-m+1}\mathbf{B}\right\},  \label{rho} \\
\mathbf{S}=& \kappa_0\mathbf{E}\times[\kappa(\mathbf{r},i\partial_t)\mathbf{B}]. \label{S}  
\end{align}}%
The conservation law (\ref{cone}) for the final expressions (\ref{rho}) and (\ref{S}) can also be verified using Maxwell's equations (\ref{max1})--(\ref{max2}). In (\ref{S}) we see that the energy flux is given by the Poynting vector $\mathbf{E}\times\mathbf{H}$ in the time domain, the obvious generalization of the non-dispersive result. The energy density (\ref{rho}), however, has no simple relation to the non-dispersive result, which only contains the first two terms in (\ref{rho}). It is a simple matter to express (\ref{rho}) and (\ref{S}) in the frequency domain.

We can use the exact energy density (\ref{rho}) to derive the standard textbook result~\cite{jac,LLcm} for the time-averaged energy density of a monochromatic wave in a dispersive medium. Such a wave has an electric field
\begin{equation}  \label{Emono}
\mathbf{E}(\mathbf{r},t)=\frac{1}{2}\left(\mathbf{E}_0 (\mathbf{r})e^{-i\omega_0 t}+\text{c.c}\right), \\
\end{equation}
and the $\mathbf{B}$ field is of the same form. When the monochromatic $\mathbf{E}$ and $\mathbf{B}$ fields are substituted into (\ref{rho}) and a time average is taken, all $t$-dependent terms vanish. The series in (\ref{rho}) containing the $\mathbf{E}$ field is then
\begin{gather}
\frac{\varepsilon_0}{4}\sum_{n=1}^\infty\sum_{m=1}^{2n}\varepsilon_{2n}(\mathbf{r})\omega_0^{2n}\left|\mathbf{E}_0\right|^2=\frac{\varepsilon_0}{4}\sum_{n=1}^\infty 2n\,\varepsilon_{2n}(\mathbf{r})\omega_0^{2n}\left|\mathbf{E}_0\right|^2  \nonumber \\
=\frac{\varepsilon_0}{4} \omega_0\frac{d\varepsilon(\mathbf{r}, \omega_0)}{d \omega_0}\left|\mathbf{E}_0\right|^2. 
\end{gather}
The series in (\ref{rho}) containing the $\mathbf{B}$ field undergoes a similar simplification and the time-averaged monochromatic energy density $\bar{\rho}_{\text{mono}}$ can be written
\begin{equation}  \label{brill}
\bar{\rho}_{\text{mono}}=\frac{\varepsilon_0}{4}\frac{d[\omega_0\varepsilon(\mathbf{r}, \omega_0)]}{d \omega_0}\left|\mathbf{E}_0\right|^2+\frac{\mu_0}{4}\frac{d[\omega_0\mu(\mathbf{r}, \omega_0)]}{d \omega_0}\left|\mathbf{H}_0\right|^2,
\end{equation}
which is Brillouin's formula~\cite{jac,LLcm}. For homogeneous media the monochromatic wave is a plane wave and it is easy to show, by dividing the time-averaged energy flow $\bar{\mathbf{S}}$ by $\bar{\rho}$, that the electromagnetic energy of the plane wave moves through the medium at the group velocity $c/[d (\omega_0\sqrt{\varepsilon(\omega_0)\mu(\omega_0)})/d\omega_0]$. It is of course expected that the energy of a monochromatic wave in a lossless, homogeneous medium should move at the group velocity, but here we have derived this fact for electromagnetism in lossless media with arbitrary dispersion.

In the case of fields that are nearly monochromatic, Brillouin's expression (\ref{brill}) can serve as an approximation of the time-averaged energy density, but only if $\varepsilon(\mathbf{r}, \omega)$ and $\mu(\mathbf{r}, \omega)$ do not vary significantly over the range of frequency components in the fields~\cite{jac}. As was recently pointed out in~\cite{ros10}, Brillouin's formula is also exact for time-averaged fields whose frequency components are uncorrelated, the important example being thermal radiation. But the exact energy density for all fields in dispersive, lossless media is given by (\ref{rho}).

\begin{figure}
\begin{center}
\includegraphics[width=8cm]{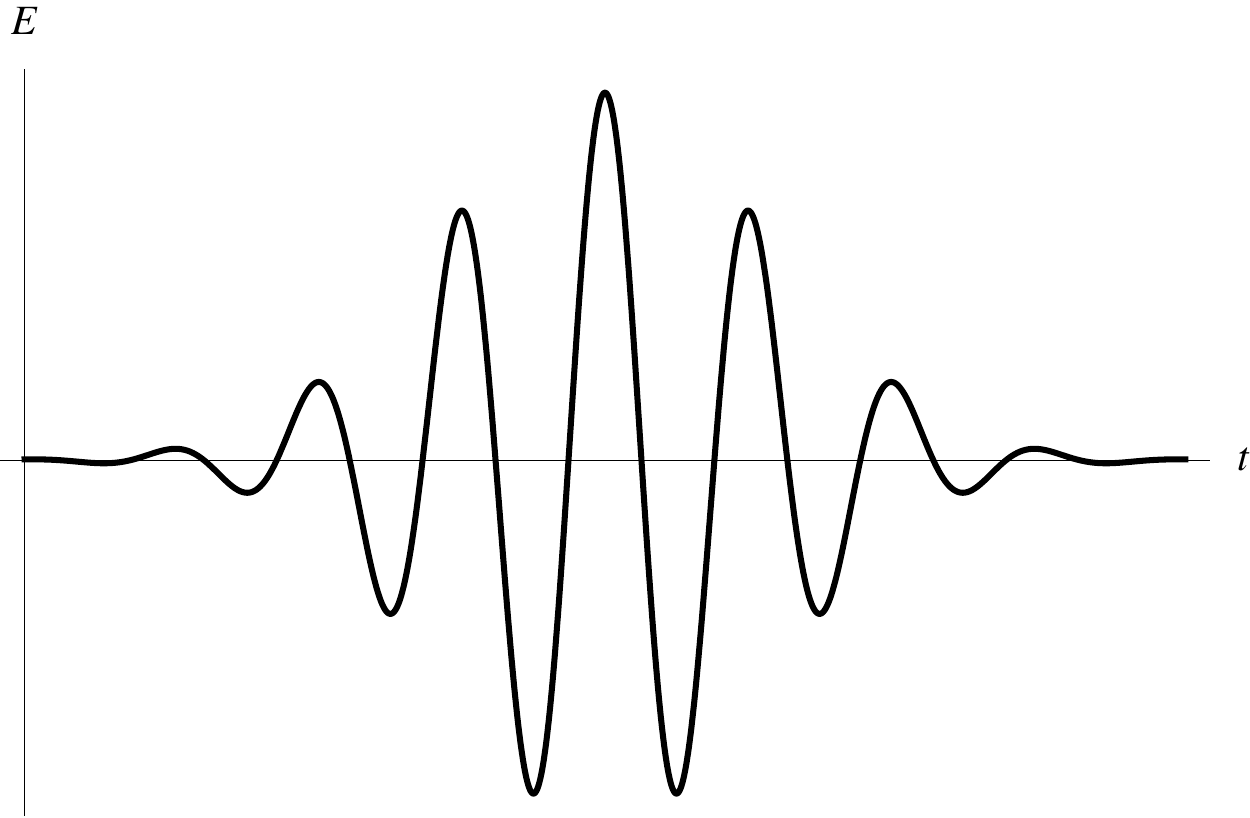}
\includegraphics[width=8cm]{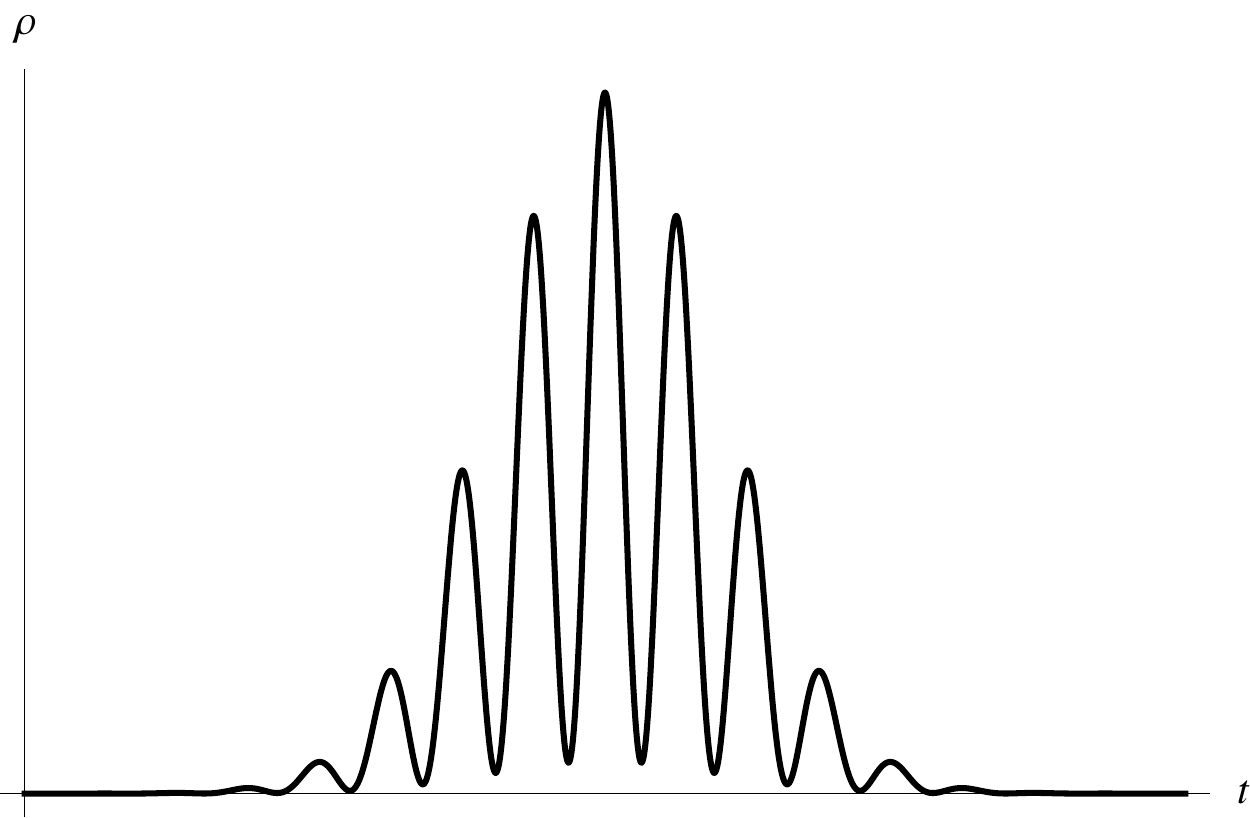}
\caption{The temporal profile of the electric field (top) and energy density (bottom) of a gaussian pulse in a medium with relative permittivity (\ref{eposc}) and $\mu=1$. Note that the energy density is not zero at all the local minima; this feature is due solely to the complicated electric-field series in (\ref{rho}).} \label{fig:en}
\end{center} 
\end{figure}

As mentioned in Section~\ref{sec:act}, the series representations (\ref{matseries}) on which our results are based will in practice be numerical fits to the measured dispersion data of the material in question, in the frequency range of interest. Clearly, a finite number of terms in the series will be sufficient for an accurate treatment of the dispersion, and the forgoing results hold in this case where all series terminate. As a theory exercise, however, we now consider an example of an infinite series expansion of the form (\ref{matseries}) for the permittivity, and verify that the expression (\ref{rho}) for the energy density is a well defined quantity. We take the standard, damped harmonic oscillator model of a homogeneous permittivity~\cite{jac}. At frequencies well below the resonant frequency $\omega_0$ of the oscillator, the imaginary part of the relative permittivity is negligible; the real part is
{\allowdisplaybreaks
\begin{align}
\varepsilon(\omega)=&1-\frac{\Omega(\omega^2-\omega_0^2)}{(\omega^2-\omega_0^2)^2+\gamma^2\omega^2}  \nonumber \\
=&1+\sum_{n=0}^{\infty}C_{2n}\omega^{2n},   \qquad \omega<\omega_0,   \label{eposc} \\
C_{2n}=&\frac{\Omega}{\omega_0^2\omega^{2n}}\sum_{m=0}^n(-1)^m
\left(
\begin{array}{c}
  n+m  \\
  2m
\end{array}
\right)
\left(\frac{\gamma^2}{\omega_0^2}\right)^m.
\end{align}}
The series expansion (\ref{eposc}) converges for $ \omega<\omega_0$ and we consider a light beam in this material with a spectrum lying far enough below $\omega_0$ for losses to be negligible. The evolution of the light beam in this dispersive medium can of course only be computed numerically, but we consider an initial, input gaussian pulse with central frequency $\omega_0/2$ and calculate its initial energy distribution. As is clear from (\ref{rho}), the effect of dispersion on the initial energy density will be seen in the temporal profile of the pulse, rather than the spatial profile, so we consider a fixed point $\mathbf{r}=0$, which can be viewed as a boundary  through which the pulse enters the medium. We take units with $c=1$ and choose the values $\omega_0=10$, $\Omega=30$ and $\gamma=1$ in the relative permittivity (\ref{eposc}). The input electric field of the gaussian pulse centered on $\omega=\omega_0/2$ at the boundary $\mathbf{r}=0$ is taken as
\begin{equation}
\mathbf{E}(t)|_{\mathbf{r}=0}=e^{-t^2/4}\cos(\omega_0t/2)
\end{equation}
and is plotted in Fig.~\ref{fig:en}. We can numerically compute the input magnetic field $\mathbf{B}(t)|_{\mathbf{r}=0}$ by transforming to the frequency domain and use of the dispersion relation $k=\sqrt{\varepsilon}\omega/c$ and the Maxwell equation $i\omega\tilde{\mathbf{B}}=\nabla\times\tilde{\mathbf{E}}$. To avoid inessential complications we ignore the transverse spatial profile of the pulse and consider a one-dimensional propagation (an actual one-dimensional propagation in a waveguide such as an optical fiber will involve an effective dispersion different from (\ref{eposc}), but our considerations here are purely for demonstration purposes). Using $\mathbf{E}(t)|_{\mathbf{r}=0}$ and  $\mathbf{B}(t)|_{\mathbf{r}=0}$ we compute the input temporal profile  $\rho(t)|_{\mathbf{r}=0}$ of the energy density (\ref{rho}) (with $\mu=1/\kappa=1$) and the result is shown in Fig.~\ref{fig:en}. Note that no cycle-averaging has been performed; the exact energy density (\ref{rho}) is not a cycle-averaged quantity. As well as the quantitative change in the energy density caused by the dispersion, there is a qualitative difference, visible in Fig.~\ref{fig:en}, compared to the same pulse in a non-dispersve medium. The local minima of the energy density in Fig.~\ref{fig:en} occur at nodes of the electric (and magnetic) field, but it is clearly seen that the energy density does not drop to zero at many of these local minima. This is purely an effect of the dispersion; in a non-dispersive medium these local minima are zeros of the energy density. In fact this effect of dispersion is due solely to the complicated electric-field series in (\ref{rho}). The familiar $\mathbf{E}\cdot\mathbf{D}$ term in (\ref{rho}) does not contribute to this feature, which can only be seen using the exact result (\ref{rho}).

\section{Momentum}  \label{sec:mom}
Turning to the electromagnetic momentum, we note that momentum will be conserved only if the medium is homogeneous, so that Noether's theorem applies to spatial translations. It is instructive however to retain $\varepsilon$ and $\kappa$ as functions of position and thereby see how the momentum conservation law fails to hold in the inhomogeneous case. To extract the conservation law associated with spatial translation invariance we make the active infinitesimal translation $\phi(\mathbf{r},t)\rightarrow\phi(\mathbf{r}+\bm{\eta}(\mathbf{r},t),t)$, $\mathbf{A}(\mathbf{r},t)\rightarrow\mathbf{A} (\mathbf{r}+\bm{\eta}(\mathbf{r},t),t)$ in the action (\ref{act}). Noether's theorem~\cite{wei} shows that when the resulting change in the action is written in the form
\begin{equation} \label{actvarr}
\delta\mathcal{S}=-\int d^4x\left(P_i\,\partial_t\eta^i+\sigma_i^{\ j}\,\nabla_j\eta^i\right),
\end{equation}
then (in homogeneous media) $P_i$ and $\sigma_i^{\ j}$ obey the conservation law
\begin{equation} \label{conm}
\partial_t P_i+ \nabla_j\sigma_i^{\ j}=0.
\end{equation}
In this way we find the electromagnetic momentum density $\mathbf{P}$ and stress tensor $\sigma_i^{\ j}$. We will carry out this procedure with $\varepsilon$ and $\kappa$ varying in space; then (\ref{conm}) will fail to hold because extra terms will appear in this equation that contain spatial derivatives of $\varepsilon$ and $\kappa$. Again, use must be made of the identities (\ref{sum1}) and (\ref{sum2}) to achieve the form (\ref{actvarr}) and the result is
{\allowdisplaybreaks
\begin{align}
 P_i=&\frac{\kappa_0}{2}\left[\frac{2}{c^2}\nabla_iA^j\varepsilon(\mathbf{r},i\partial_t)E_j\right. \nonumber \\
&+\frac{1}{c^2}\sum_{n=1}^\infty\sum_{m=1}^{2n}(-1)^{n+m}\varepsilon_{2n}(\mathbf{r})\partial_t^{m-1}E_j\partial_t^{2n-m}\nabla_iE^j  \nonumber \\
&-\left.\sum_{n=1}^\infty\sum_{m=1}^{2n}(-1)^{n+m}\kappa_{2n}(\mathbf{r})\partial_t^{m-1}B_j\partial_t^{2n-m}\nabla_iB^j \right],  \label{P1} \\
\sigma_i^{\ j}=&\, \mathcal{L}\,\delta_i^{\ j}+\varepsilon_0\nabla_i\phi\,\varepsilon(\mathbf{r},i\partial_t)E^j   \nonumber \\
&+\kappa_0\nabla_iA_k\kappa(\mathbf{r},i\partial_t)[\nabla^jA^k-\nabla^kA^j], \label{stress1}  
\end{align}}%
where $\mathcal{L}$ in (\ref{stress1}) is the Lagrangian density, i.e.\ the integrand in the action (\ref{act}). To obtain gauge-invariant expressions for the momentum density and stress tensor, we note that the quantities
\begin{gather}
f^{j\ t}_{\ \,i}:= -\varepsilon_0A_i\varepsilon(\mathbf{r},i\partial_t)E^j =:-f^{t\ j}_{\ \,i},  \\
f^{k\ j}_{\ \,i}:=-\kappa_0A_i\kappa(\mathbf{r},i\partial_t)(\nabla^jA^k-\nabla^kA^j),
\end{gather}
identically satisfy
\begin{equation}  \label{fid2}
\partial_t\nabla_jf^{j\ t}_{\ \,i}+\nabla_j(\partial_tf^{t\ j}_{\ \,i}+\nabla_kf^{k\ j}_{\ \,i})=0.
\end{equation}
Thus, addition of $\nabla_jf^{j\ t}_{\ \,i}$ to $P_i$, and of $\partial_tf^{t\ j}_{\ \,i}+\nabla_kf^{k\ j}_{\ \,i}$ to $\sigma_i^{\ j}$, does not affect the momentum conservation law (\ref{conm}). After these additions and use of Maxwell's equations (\ref{max1})--(\ref{max2}), the momentum density and stress tensor are gauge invariant and are given by
{\allowdisplaybreaks
\begin{align}
 P_i=&\, \varepsilon_0\epsilon_{ijk}[\varepsilon(\mathbf{r},i\partial_t)E^j]B^k \nonumber \\
&+\frac{\varepsilon_0}{2}\sum_{n=1}^\infty\sum_{m=1}^{2n}(-1)^{n+m}\varepsilon_{2n}(\mathbf{r})\partial_t^{m-1}E_j\partial_t^{2n-m}\nabla_iE^j  \nonumber \\
&-\frac{\kappa_0}{2}\sum_{n=1}^\infty\sum_{m=1}^{2n}(-1)^{n+m}\kappa_{2n}(\mathbf{r})\partial_t^{m-1}B_j\partial_t^{2n-m}\nabla_iB^j,  \label{P} \\
\sigma_i^{\ j}=&-\varepsilon_0E_i\varepsilon(\mathbf{r},i\partial_t)E^j-\kappa_0[\kappa(\mathbf{r},i\partial_t)B_i]B^j   \nonumber \\
&+\frac{1}{2}\delta_i^{\ j}\left[\varepsilon_0E_k\varepsilon(\mathbf{r},i\partial_t)E^k+\kappa_0B_k\kappa(\mathbf{r},i\partial_t)B^k \right], \label{stress}  
\end{align}}%
where $\epsilon_{ijk}$ is the (completely antisymmetric) Levi-Civita tensor. It is straightforward to verify that, when the fields satisfy Maxwell's equations (\ref{max1})--(\ref{max2}), the  momentum density (\ref{P}) and stress tensor (\ref{stress}) satisfy
\begin{align} 
\partial_t P_i+ \nabla_j\sigma_i^{\ j}=&\frac{\varepsilon_0}{2}E_j[\nabla_i\varepsilon(\mathbf{r},i\partial_t)]E^j \nonumber \\
&-\frac{\kappa_0}{2}B_j[\nabla_i\kappa(\mathbf{r},i\partial_t)]B^j,
\label{conminhom}
\end{align}
so that the momentum conservation law (\ref{conm}) indeed holds for homogeneous media. The derivation of (\ref{conminhom}) in the non-dispersive case is familiar from the textbooks~\cite{str}. 

Note that the electromagnetic stress tensor (\ref{stress}) is here defined so that it is the spatial part of the electromagnetic energy-momentum tensor; a widespread convention for the stress tensor differs from this by a minus sign~\cite{jac}. The stress tensor (\ref{stress}) is the obvious generalization to the dispersive case of the non-dispersive result; the fact that this expression describes the stress tensor in dispersive media is deduced in~\cite{LLcm} from completely different considerations.
 
For the monochromatic wave (\ref{Emono}), the series in the momentum density (\ref{P}) containing the $\mathbf{E}$ field is, after a time averaging,
\begin{align} 
\frac{\varepsilon_0}{8}\sum_{n=1}^\infty\sum_{m=1}^{2n}i\varepsilon_{2n}\omega_0^{2n-1}\left(E_{0j}\nabla_i E^{*j}_0-E^*_{0j}\nabla_i E^j_0\right)  \nonumber \\
=-\frac{\varepsilon_0}{4}\frac{d\varepsilon(\omega_0)}{d\omega_0}\text{Im}(E_{0j}\nabla_i E^{*j}_0).  
\end{align}
There is a similar result for the series in (32) containing the magnetic field, and the complete result for the time-averaged monochromatic momentum density $\bar{\mathbf{P}}$ in a homogeneous medium is
\begin{align}   
\bar{\mathbf{P}}=& \frac{\varepsilon_0}{2}\varepsilon(\omega_0)\text{Re}(\mathbf{E}_0\times\mathbf{B}^*_0)-\frac{\varepsilon_0}{4}\frac{d\varepsilon(\omega_0)}{d\omega_0}\text{Im}(E_{0i}\nabla E^{*i}_0)  \nonumber \\
&+\frac{\kappa_0}{4}\frac{d\kappa(\omega_0)}{d\omega_0}\text{Im}(B_{0i}\nabla B^{*i}_0).  \label{Pmono}
\end{align}
The Minkowski momentum density $\mathbf{D}\times\mathbf{B}$ would give only the first term in (\ref{Pmono}). With an additional restriction to a single plane wave of frequency $\omega_0$ in the homogeneous medium, i.e.
\begin{gather}  
\mathbf{E}_0(\mathbf{r})=\bm{\mathcal{E}}_0 e^{i\mathbf{k}\cdot\mathbf{r}}, \qquad \mathbf{B}_0(\mathbf{r})=\frac{1}{\omega_0}\mathbf{k}\times\mathbf{E}_0(\mathbf{r}), \\
|\mathbf{k}|=\frac{\omega_0}{c}\sqrt{\varepsilon(\omega_0)\mu(\omega_0)},
\end{gather}  
the expression (\ref{Pmono}) reduces to
\begin{gather}   
\bar{\mathbf{P}}=\frac{\varepsilon_0}{2c}\varepsilon(\omega_0)n_g(\omega_0)\mathbf{\hat{k}}|\bm{\mathcal{E}}_0|^2,  \label{Pmono2} \\
 n_g(\omega_0)=n_p(\omega_0)+\omega_0\frac{dn_p(\omega_0)}{d\omega_0},
\end{gather} 
where $n_g(\omega_0)$ is the group index, $n_p(\omega_0)=\sqrt{\varepsilon(\omega_0)\mu(\omega_0)}$ is the phase index and $\mathbf{\hat{k}}=\mathbf{k}/|\mathbf{k}|$. Using the quantum prescription (for $\mu=1$) in~\cite{gar} corresponds to taking $|\bm{\mathcal{E}}_0|^2\to 2\hbar\omega_0/[\varepsilon_0n_p(\omega_0)n_g(\omega_0)]$ for a single photon; this gives from (\ref{Pmono2}) a single photon momentum that is $n_p(\omega_0)$ times the vacuum value, in agreement with the  canonical momentum in~\cite{bar10b}. We note that the Minkowski expression (the first term in (\ref{Pmono})) would give a correction factor of $n^2_p(\omega_0)/n_g(\omega_0)$ for the single photon~\cite{gar}, whereas the factor $n_p(\omega_0)$ has been obtained here from the general momentum density (32).

\begin{figure}
\begin{center}
\includegraphics[width=8cm]{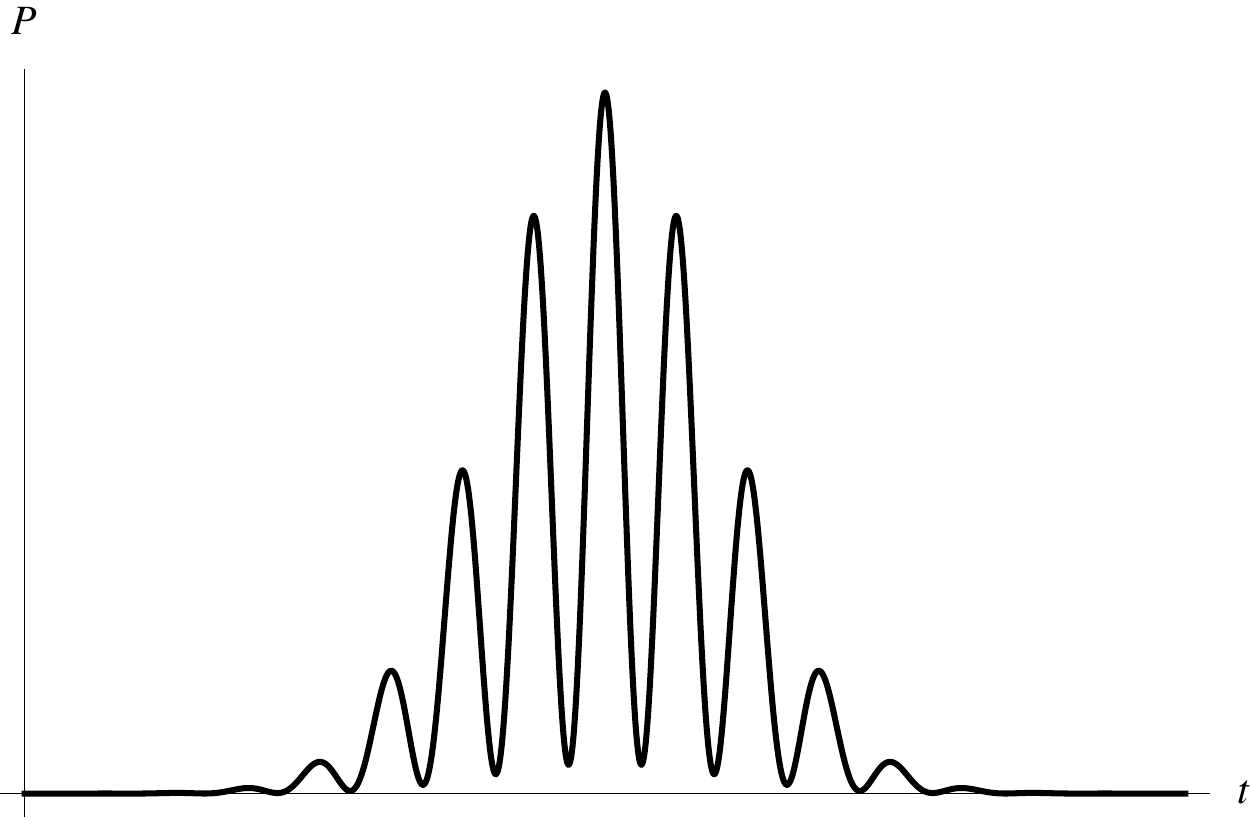}
\caption{The temporal profile of the momentum density of a gaussian pulse in a dispersive medium. The pulse and dispersion are the same as those described at the end of Section~\ref{sec:en} and used for the energy density in Fig.~\ref{fig:en}. Note that the momentum density is not zero at local minima; this feature is due solely to the complicated electric-field series in (\ref{P}).} \label{fig:mom}
\end{center}
\end{figure}

At the end of Section~\ref{sec:en} we showed an example of the electromagnetic energy density for a gaussian pulse in a dispersive medium. For the same pulse and medium we use (\ref{P}) to compute the input temporal profile of the momentum density, where the final series in the magnetic field in (\ref{P}) is absent for this example. The spatial derivative of the electric field in (\ref{P}) is computed by transforming to the frequency domain and using the dispersion relation $k=\sqrt{\varepsilon}\omega/c$ (as discussed in Section~\ref{sec:en}, we assume one-dimensional propagation). The profile for the momentum density is shown in Fig.~\ref{fig:mom} and is seen to have the same shape as the energy density in Fig.~\ref{fig:en}. Similar to what was found for the energy density, the fact that the momentum density does not fall to zero at local minima is due solely to the complicated electric-field series in (\ref{P}); this feature can only be seen with the exact expression (\ref{P}) and is absent from the Minkowski contribution $\mathbf{D}\times\mathbf{B}$ in (\ref{P}).

Electromagnetic angular momentum in dispersive media can be derived in a similar manner, though this will not be done here. The angular momentum density is not to be constructed from the energy-momentum tensor but rather derived from the invariance of the action (\ref{act}) under active rotations (for rotationally symmetric media).

\section{Conclusions}
We have derived the exact conserved energy-momentum tensor of light propagating in lossless, dispersive, homogeneous media. Electromagnetic energy is also conserved when the medium is inhomogeneous and this case has been included. The energy flux and stress tensor in dispersive media have the same general form as in the non-dispersive case, with the permittivity and permeability becoming derivative operators in the time domain. In contrast, the energy density and momentum density for arbitrary fields have no simple relation to the non-dispersive results. For time-averaged monochromatic waves the Brillouin energy density is recovered, and the momentum density also takes a simple form (equation (\ref{Pmono})). As stressed in the Introduction, the results do not address the question of momentum balance in light-matter interactions, which requires an analysis of momentum transfer between light and matter~\cite{bre79,leo06,pfe07,bar10a,bar10b}. The conserved energy-momentum tensor of light in lossless media is a well-defined quantity with its own experimental significance~\cite{phi08}; it has been derived here in complete generality within the framework of macroscopic electromagnetism.

\section*{Acknowledgements}
I thank Oliver Allanson for pointing out an error in the original monochromatic momentum result. This research is supported by the Scottish Government and the Royal Society of Edinburgh.

\end{document}